\documentstyle[graphicx,12pt]{article}
\begin{document}

\small
\hoffset=-1truecm
\voffset=-2truecm
\title{\bf The Casimir force on a piston in the spacetime with extra compactified dimensions}
\author{Hongbo Cheng\footnote {E-mail address:
hbcheng@public4.sta.net.cn}\\
Department of Physics, East China University of Science and
Technology,\\ Shanghai 200237, China}

\date{}
\maketitle

\begin{abstract}
A one-dimensional Casimir piston for massless scalar fields
obeying Dirichlet boundary conditions in high-dimensional
spacetimes within the frame of Kaluza-Klein theory is analyzed. We
derive and calculate the exact expression for the Casimir force on
the piston. We also compute the Casimir force in the limit that
one outer plate is moved to the extremely distant place to show
that the reduced force is associated with the properties of
additional spatial dimensions. The more dimensionality the
spacetime has, the stronger the extra-dimension influence is. The
Casimir force for the piston in the model excluding one plate
under the background with extra compactified dimensions always
keeps attractive. Further we find that when the limit is taken the
Casimir force between one plate and the piston will change to be
the same form as the corresponding force for the standard system
consisting of two parallel plates in the four-dimensional
spacetimes if the ratio of the plate-piston distance and extra
dimensions size is large enough.
\end{abstract}

\vspace{6cm}
\hspace{1cm}
PACS number(s): 03.70.+k; 11.10Kk

\newpage

In 1948 a remarkable macroscopic quantum effect describing the
attractive force between two conducting and neutral parallel
plates was predicted by Casimir [1]. The Casimir effect appears
due to the disturbance of the vacuum of the electromagnetic field
induced by the presence of boundary. Twenty years later Boyer
researched on the Casimir effect for a conducting spherical shell
to find that this kind of Casimir force is repulsive [2]. This
effect is more complicated than we thought. Afterwards more
efforts have been paid for the problem and related topics. All
results brought attention to the fact that whether the Casimir
force is attractive or repulsive depends on the geometry of the
configuration strongly [3]. However, there are several reasons to
be suspicious of the analysis of the Casimir effect problems.
Maybe their results are not perfect. For example, we always
investigated a massless scalar field in a confined region such as
parallel plates, rectangular box and so on to find the vacuum
energy while we let the field satisfy the Dirichlet boundary
conditions at the borders of the region [3-8]. Having regularized
the vacuum energy, we obtain the Casimir energy. Certainly the
Casimir force can be received by means of derivative of Casimir
energy with respect to the distance between two edges. Here it
should be pointed out that these former considerations on the
topic have not involved the contribution to the vacuum energy from
the area outside the confined region which depends on its
dimensions while we discard the divergent terms related to the
boundary also depending on the geometry and dimensions during the
regularization process. In order to ignore the flaws mentioned
above, a slightly different model called piston was put forward
[9]. The system is a single rectangular box with dimensions
$L\times b$ divided into two parts with dimensions $a\times b$ and
$(L-a)\times b$ respectively by a piston which is an idealized
plate that is free to move along a rectangular shaft. In ref. [9]
the author calculated the Casimir force on a two-dimensional
piston as a consequence of fluctuations of a scalar field obeying
Dirichlet boundary conditions on all surfaces and found that the
force on the piston is always attractive as $L$ goes to the
infinity, regardless of the ratio of the two sides. Immediately
the issue attracted more attention. The Casimir force acting on a
conducting piston with arbitrary cross section always keeps
attractive although the existence of the walls weaken the force
[10]. The three-dimensional Casimir piston for massless scalar
fields obeying Dirichlet boundary conditions was also explored,
and it was found that the total Casimir force is negative no
matter how long the lengths of sides are [11]. In addition in the
case of various boundary conditions the Casimir force on a piston
may be repulsive [12, 13]. In a word the Casimir piston is a new
important model revealing its own distinct effects and can be used
to explore the related topics. This model is also simpler to be
constructed as a device from the experimental point of view.

The model of higher-dimensional spacetime is a powerful ingredient
to be needed to unify the interactions in nature. More than 80
years ago Kaluza and Klein put forward the issue that our universe
has more than four dimensions [14, 15]. The Kaluza-Klein theory
introduced an extra compactified dimension to unify gravity and
classical electrodynamics in our world. The theory has been
generalized and developed greatly. Recently the quantum gravity
such as string theory or brane-world scenario is developed to
reconcile the quantum mechanics and gravity with the help of
introducing seven additional spatial dimensions. In
Randall-Sundrum model the matter fields may be localized on a
four-dimensional brane considered as our real universe, and only
gravitons can propagate in the extra space transverse to the brane
[16, 17]. In addition, although the order of the compactification
scale of the additional dimensions has not been confirmed and are
also of considerable interest recently, larger extra dimensions
were invoked in order to provide a breakthrough of hierarchy
problem in some approaches [18-20]. Research on higher-dimensional
spacetime is valuable and become a focus in the physical
community, therefore the theory needs to be explored deeply,
extensively and in various directions.

Since the higher-dimensional spacetime described by Kaluza-Klein
theory is important and indispensable, it is crucial to discuss
several models including the Casimir effect problem in this
background. The precision of the measurement has been greatly
improved practically [21-24], leading the Casimir effect to be
remarkable observable and trustworthy consequence of the existence
of quantum fluctuations and to become a powerful tool for the
topics on the model of Universe with more than four dimensions. It
must be emphasized that the attractive Casimir force between the
parallel plates vanishes when the plate gap is very large and no
repulsive force appears according to the experimental results.
Some topics were examined in the context of Kaluza-Klein theory.
As the first step of generalization to investigate the
higher-dimensional spacetimes, we show analytically that the
extra-dimension corrections to the Casimir effect for a
rectangular cavity in the presence of a compactified universal
extra dimension are very manifest [25]. The Casimir effect for
parallel plates in the spacetime with extra compactified
dimensions was also studied [26-29]. We prove rigorously that
there must appear repulsive Casimir force between the parallel
plates when the plates distance is sufficiently large in the
spacetime with compatified additional dimensions, and the higher
the dimensionality is, the greater the repulsive force is. It
should be pointed out that the value of the repulsive Casimir
force which is obtained theoretically is within the experimental
reach. Therefore the results obtained in the context of
Kaluza-Klein theory conflict with the experimental phenomena
mentioned above [27-29], which means that the model of
higher-dimensional spaetime with extra compactified spatial
dimensions needs further research.

It is necessary and significant to study the force-on-the-piston
problem in a higher-dimensional spacetime within the frame of
Kaluza-Klein theory. We wonder how the influence from extra
dimensions on the Casimir effect of the piston. This problem, to
our knowledge, has not been examined. For simplicity and
comparison to the conclusion of standard parallel-plates system,
the model of one-dimensional piston is chosen. The main purpose of
this paper is to study the Casimir effect for the system
consisting of three parallel plates in the Universe with $d$
compactified spatial dimensions. We obtain the expression of force
by means of the derfferential of the total vacuum energy including
the contribution outside the three-parallel-plate device with
respect to the distance between two plates in the system. We
regularize the force to obtain the Casimir force on the piston
when one outer plate is moved to the remote place. We focus on the
influence of dimensionality of the spacetime on the Casimir force
between one plate and a piston and compare our results with the
models like one-dimensional piston or parallel plates in the
four-dimensional spacetime. Our discussions and conclusions are
emphasized in the end.

In a higher-dimensional spacetime, we start to consider the
massless scalar fields obeying Dirichlet boundary conditions
within a one-dimensional piston. As a piston, one plate is
inserted into a system consisting of two parallel plates. The
piston is parallel to the plates and divides the
parallel-plate-system into two parts labeled by $A$ and $B$
respectively. In part $A$ the distance between the left plate and
the piston is $a$, and the distance between the piston and the
right plate in part $B$, the remains of the separation of two
plates, is certainly $L-a$, which means that $L$ denotes the whole
plates gap. The total vacuum energy for the three-parallel-plate
system described above can be written as the sum of three terms,

\begin{equation}
E=E^{A}(a)+E^{B}(L-a)+E^{out}
\end{equation}

\noindent where $E^{A}(a)$ and $E^{B}(L-a)$ represent the energy
of part $A$ and $B$ respectively, and the terms depend on their
each size in parts. $E^{out}$ describes the vacuum energy outside
the system and is independent of characters inside the system.
Having regularized the total energy density, we obtain the Casimir
energy density,

\begin{equation}
E_{C}=E^{A}_{R}(a)+E^{B}_{A}(L-a)+E^{out}_{R}
\end{equation}

\noindent where $E^{A}_{R}(a)$, $E^{B}_{R}(L-a)$ and $E^{out}_{R}$
are finite parts of terms $E^{A}(a)$, $E^{B}(L-a)$ and $E^{out}$
in Eq.(1) respectively. In particular, it is also pointed out that
$E^{out}_{R}$ is not a function of the position of the piston, the
Casimir force on the piston is given by the derivative of the
Casimir energy with respect to the plates distance
$-\frac{\partial E_{C}}{\partial a}$ and can be written as,

\begin{equation}
F_{C}=-\frac{\partial}{\partial a}[E^{A}_{R}(a)+E^{B}_{R}(L-a)]
\end{equation}

\noindent which means that the contribution of vacuum energy from
the exterior region does not affect the Casimir force on the
piston.

Here we set out to consider the massless scalar field in the
three-parallel-plate system in the spacetime with $d$ extra
compactified dimensions in the context of Kaluza-Klein theory.
Along the additional dimensions the wave vectors of the field have
the form $k_{i}=\frac{n_{i}}{R}$, $i=1,2,\cdot\cdot\cdot,d$,
respectively, $n_{i}$ an integer. Now we choose that the extra
dimensions possess the same size as $R$. The fields satisfy the
Dirichlet condition, leading the wave vector in the directions
restricted by the plates to be $k_{n}=\frac{n\pi}{D}$, $n$ a
positive integer and $D$ the separation of the plates. Under these
conditions, the zero-point fluctuations of the fields can give
rise to observable Casimir forces among the plates.

In the case of $d$ additional compactified dimensions we find the
frequency of the vacuum fluctuation within a region confined by
two parallel plates whose separation is $D$ to be,

\begin{equation}
\omega_{\{n_{i}\}n}=\sqrt{k^{2}+\frac{n^{2}\pi^{2}}{D^{2}}
+\sum_{i=1}^{d}\frac{n_{i}^{2}}{R^{2}}}
\end{equation}

\noindent where

\begin{equation}
k^{2}=k_{1}^{2}+k_{2}^{2}
\end{equation}

\noindent $k_{1}$ and $k_{2}$ are the wave vectors in directions
of the unbound space coordinates parallel to the plates surface.
Now $\{n_{i}\}$ represents a short notation of
$n_{1},n_{2},\cdot\cdot\cdot,n_{d}$, $n_{i}$ a nonnegative
integer. According to Ref.[3, 4, 7, 30-34], therefore the total
energy density of the fields in the interior of two-parallel-plate
system reads,

\begin{eqnarray}
E(D,R)=\int d^{2}k\sum_{n=1}^{\infty}\sum_{\{n_{i}\}=0}^{\infty}
\frac{1}{2}\omega_{\{n_{i}\}n} \hspace{6.5cm}\nonumber\\
=\frac{\pi}{2}\frac{\Gamma(-\frac{3}{2})}{\Gamma(-\frac{1}{2})}
\sum_{l=0}^{d-1}\left(%
\begin{array}{c}
  d \\
  l \\
\end{array}%
\right)
E_{d-l+1}(\frac{\pi^{2}}{D^{2}},\frac{1}{R^{2}},\frac{1}{R^{2}},\cdot\cdot\cdot,
\frac{1}{R^{2}};-\frac{3}{2})+\frac{\pi^{4}}{2D^{3}}
\frac{\Gamma(-\frac{3}{2})\zeta(-3)}{\Gamma(-\frac{1}{2})}
\end{eqnarray}

\noindent in terms of the Epstein zeta function
$E_{p}(a_{1},a_{2},\cdot\cdot\cdot,a_{p};s)$ defined as,

\begin{equation}
E_{p}(a_{1},a_{2},\cdot\cdot\cdot,a_{p};s)=\sum_{\{n_{j}\}}^{\infty}
(\sum_{j=1}^{p}a_{j}n_{j}^{2})^{-s}
\end{equation}

\noindent here $\{n_{j}\}$ stands for a short notation of
$n_{1},n_{2},\cdot\cdot\cdot,n_{p}$, $n_{j}$ a positive integer.
We can regularize Eq.(6) by means of the following formula,

\begin{eqnarray}
\Gamma(-\frac{3}{2})E_{d-l+1}(\frac{\pi^{2}}{D^{2}},\frac{1}{R^{2}},
\frac{1}{R^{2}},\cdot\cdot\cdot,\frac{1}{R^{2}};-\frac{3}{2})\hspace{5.5cm}\nonumber\\
=-\frac{1}{2}\Gamma(-\frac{3}{2})E_{d-l}(1,1,\cdot\cdot\cdot,1;-\frac{3}{2})
\frac{1}{R^{3}}+\frac{1}{2\sqrt{\pi}}\Gamma(-2)E_{d-l}
(1,1,\cdot\cdot\cdot,1;-2)\frac{D}{R^{4}}\nonumber\\
+\frac{1}{R^{3}}\sum_{k=0}^{\infty}\frac{16^{-k}}{k!}(\frac{D}{R})^{-k-\frac{3}{2}}
\prod_{j=1}^{k}[16-(2j-1)^{2}]\hspace{4cm}\nonumber\\
\times\sum_{n_{1},n_{2},\cdot\cdot\cdot,n_{d-l+1}=1}
^{\infty}n_{1}^{-k-\frac{5}{2}}(n_{2}^{2}+n_{3}^{2}+\cdot\cdot\cdot
+n_{d-l+1}^{2})^{-\frac{2k-3}{4}}\hspace{1cm}\nonumber\\
\times\exp[-\frac{2D}{R}n_{1}(n_{2}^{2}
+n_{3}^{2}+\cdot\cdot\cdot+n_{d-l+1}^{2})^{\frac{1}{2}}]\hspace{3cm}
\end{eqnarray}

\noindent to obtain the Casimir energy density of a system
consisting of two parallel plates in the spacetime with $d$ extra
compactified spatial dimensions.

In the context of Kaluza-Klein theory we return to discuss the new
system where a piston is also a plate localizing parallelly
between two parallel plates mentioned above. Choosing the variable
$D=a$ in Eq.(6) and (8), we have the vacuum energy density for
part $A$ of the system containing one plate and the piston with
distance $a$ as follow,

\begin{equation}
E^{A}(a,R)=\frac{\pi}{2}\frac{\Gamma(-\frac{3}{2})}{\Gamma(-\frac{1}{2})}
\sum_{l=0}^{d-1}\left(%
\begin{array}{c}
  d \\
  l \\
\end{array}%
\right)
E_{d-l+1}(\frac{\pi^{2}}{a^{2}},\frac{1}{R^{2}},\frac{1}{R^{2}},
\cdot\cdot\cdot,\frac{1}{R^{2}};-\frac{3}{2})+\frac{\pi^{4}}{2a^{3}}
\frac{\Gamma(-\frac{3}{2})\zeta(-3)}{\Gamma(-\frac{1}{2})}
\end{equation}

\noindent Similarly the vacuum energy density for the remains of
the system labeled $B$ with plates distance $L-a$ by replacing the
variable $D$ with $L-a$ in Eq.(6) is,

\begin{eqnarray}
E^{A}(L-a,R)=\frac{\pi}{2}\frac{\Gamma(-\frac{3}{2})}{\Gamma(-\frac{1}{2})}
\sum_{l=0}^{d-1}\left(%
\begin{array}{c}
  d \\
  l \\
\end{array}%
\right)
E_{d-l+1}(\frac{\pi^{2}}{(L-a)^{2}},\frac{1}{R^{2}},\frac{1}{R^{2}},
\cdot\cdot\cdot,\frac{1}{R^{2}};-\frac{3}{2})\nonumber\\
+\frac{\pi^{4}}{2(L-a)^{3}}
\frac{\Gamma(-\frac{3}{2})\zeta(-3)}{\Gamma(-\frac{1}{2})}\hspace{5cm}
\end{eqnarray}

\noindent We regularize Eq.(9) and Eq.(10) and then substitute the
two regularized expressions into Eq.(3) to obtain the Casimir
force per unit area on the piston,

\begin{eqnarray}
F'_{C}=-\frac{\pi^{4}}{120}\frac{1}{a^{4}}+\frac{\pi^{4}}{120}\frac{1}{(L-a)^{4}}
\hspace{8cm}\nonumber\\
+\frac{\sqrt{\pi}}{4}\sum_{l=0}^{d-1}\left(%
\begin{array}{c}
  d \\
  l \\
\end{array}%
\right)
\{-\frac{1}{aR^{3}}\sum_{k=0}^{\infty}\frac{16^{-k}}{k!}(k+\frac{3}{2})
(\frac{a}{R})^{-k-\frac{3}{2}}\prod_{j=1}^{k}[16-(2j-1)^{2}]\nonumber\\
\times\sum_{n_{1},n_{2},\cdot\cdot\cdot,n_{d-l+1}=1}^{\infty}n_{1}^{-k-\frac{5}{2}}
(n_{2}^{2}+n_{3}^{2}+\cdot\cdot\cdot+n_{d-l+1}^{2})^{-\frac{2k-3}{4}}\nonumber\\
\times\exp[-\frac{2a}{R}n_{1}(n_{2}^{2}+n_{3}^{2}+\cdot\cdot\cdot
+n_{d-l+1}^{2})^{\frac{1}{2}}]\hspace{1.5cm}\nonumber\\
-\frac{2}{R^{4}}\sum_{k=0}^{\infty}\frac{16^{-k}}{k!}(\frac{a}{R})^{-k-\frac{3}{2}}
\prod_{j=1}^{k}[16-(2j-1)^{2}]\hspace{2cm}\nonumber\\
\times\sum_{n_{1},n_{2},\cdot\cdot\cdot,n_{d-l+1}=1}^{\infty}n_{1}^{-k-\frac{3}{2}}
(n_{2}^{2}+n_{3}^{2}+\cdot\cdot\cdot+n_{d-l+1}^{2})^{-\frac{2k-5}{4}}\nonumber\\
\times\exp[-\frac{2a}{R}n_{1}(n_{2}^{2}+n_{3}^{2}+\cdot\cdot\cdot
+n_{d-l+1}^{2})^{\frac{1}{2}}]\hspace{1.5cm}\nonumber\\
+\frac{1}{(L-a)R^{3}}\sum_{k=0}^{\infty}\frac{16^{-k}}{k!}(k+\frac{3}{2})
(\frac{L-a}{R})^{-k-\frac{3}{2}}\prod_{j=1}^{k}[16-(2j-1)^{2}]\nonumber\\
\times\sum_{n_{1},n_{2},\cdot\cdot\cdot,n_{d-l+1}=1}^{\infty}n_{1}^{-k-\frac{5}{2}}
(n_{2}^{2}+n_{3}^{2}+\cdot\cdot\cdot+n_{d-l+1}^{2})^{-\frac{2k-3}{4}}\nonumber\\
\times\exp[-\frac{2(L-a)}{R}n_{1}(n_{2}^{2}+n_{3}^{2}+\cdot\cdot\cdot
+n_{d-l+1}^{2})^{\frac{1}{2}}]\hspace{1cm}\nonumber\\
+\frac{2}{R^{4}}\sum_{k=0}^{\infty}\frac{16^{-k}}{k!}(\frac{L-a}{R})^{-k-\frac{3}{2}}
\prod_{j=1}^{k}[16-(2j-1)^{2}]\hspace{2cm}\nonumber\\
\times\sum_{n_{1},n_{2},\cdot\cdot\cdot,n_{d-l+1}=1}^{\infty}
n_{1}^{-k-\frac{3}{2}}(n_{2}^{2}+n_{3}^{2}+\cdot\cdot\cdot
+n_{d-l+1}^{2})^{-\frac{2k-5}{4}}\nonumber\\
\times\exp[-\frac{2(L-a)}{R}n_{1}(n_{2}^{2}+n_{3}^{2}+\cdot\cdot\cdot
+n_{d-l+1}^{2})^{\frac{1}{2}}]\}\hspace{1cm}
\end{eqnarray}

\noindent which has corrections from extra dimensions. Further we
take the limit $L\rightarrow\infty$ which means that the right
plate in part $B$ is moved to a very distant place, then we obtain
the following expression for the Casimir force per unit area on
the piston,

\begin{eqnarray}
F_{C}=\lim_{L\rightarrow\infty}F'_{C}\hspace{1cm}\nonumber\\
=-\frac{\pi^{4}}{120}\frac{1}{a^{4}}+\frac{C_{d}(\mu)}{R^{4}}
\end{eqnarray}

\noindent where the correction function $C_{d}(\mu)$ is defined
as,

\begin{eqnarray}
C_{d}(\mu)=\frac{\sqrt{\pi}}{4}\sum_{l=0}^{d-1}\left(%
\begin{array}{c}
  d \\
  l \\
\end{array}%
\right) \{-\frac{1}{\mu}\sum_{k=0}^{\infty}\frac{16^{-k}}{k!}
(k+\frac{3}{2})\mu^{-k-\frac{3}{2}}\prod_{j=1}^{k}[16-(2j-1)^{2}]\nonumber\\
\times\sum_{n_{1},n_{2},\cdot\cdot\cdot,n_{d-l+1}=1}^{\infty}
n_{1}^{-k-\frac{5}{2}}(n_{2}^{2}+n_{3}^{2}+\cdot\cdot\cdot
+n_{d-l+1}^{2})^{-\frac{2k-3}{4}}\nonumber\\
\times\exp[-2\mu n_{1}(n_{2}^{2}+n_{3}^{2}+\cdot\cdot\cdot
+n_{d-l+1}^{2})^{\frac{1}{2}}]\hspace{2cm}\nonumber\\
-2\sum_{k=0}^{\infty}\frac{16^{-k}}{k!}\mu^{-k-\frac{3}{2}}
\prod_{j=1}^{k}[16-(2j-1)^{2}]\hspace{3cm}\nonumber\\
\times\sum_{n_{1},n_{2},\cdot\cdot\cdot,n_{d-l+1}=1}^{\infty}
n_{1}^{-k-\frac{3}{2}}(n_{2}^{2}+n_{3}^{2}+\cdot\cdot\cdot
+n_{d-l+1}^{2})^{-\frac{2k-5}{4}}\nonumber\\
\times\exp[-2\mu n_{1}(n_{2}^{2}+n_{3}^{2}+\cdot\cdot\cdot
+n_{d-l+1}^{2})^{\frac{1}{2}}]\}\hspace{2cm}
\end{eqnarray}

\noindent and

\begin{equation}
\mu=\frac{a}{R}
\end{equation}

\noindent We discover that the first term in Eq.(12) is the same
as Casimir pressure of standard model of parallel plates involving
massless scalar fields satisfying the Dirichlet conditions in the
four-dimensional spacetimes, meaning that the Casimir force per
unit area between the piston and one plate after the other ones
has been moved away in the background with extra compactified
dimensions is just the original result plus the deviation from the
additional dimensions. It is necessary to explore the correction
functions in detail. When the gap between the remain plate and the
piston is larger enough than the size of extra dimensions, the
correction functions approach to the zero,

\begin{equation}
\lim_{\mu\rightarrow\infty}C_{d}(\mu)=0
\end{equation}

\noindent and right now the Casimir force per unit area on the
piston will return to the ones of two parallel plates in the
four-dimensional spacetime. We have to perform the burden and
surprisingly difficult calculation to scrutinize the Casimir force
on the piston quantatively, then the correction functions
$C_{d}(\mu)$ under the limiting $L\rightarrow\infty$ in the
spacetime with extra dimensions are depicted in Figure 1 and
Figure 2. Certainly the Casimir force on the piston related to the
ratio $\mu=\frac{a}{R}$ has been displayed in the same case.
According to Eq. (12), we just need to focus on the correction
functions during the process of research on the Casimir force on
the piston. The functions depend on the dimensionality of
spacetime and the ratio of plates distance and the size of extra
dimensions. The more dimensions the spacetime has, the greater the
absolute value of the correction function is, which means that
there will appear stronger influence on the Casimir force on the
piston in the higher-dimensional spacetime. The expression
$C_{d}(\mu)$ is also a function of ratio denoted in Eq.(14). When
the ratio increases, the absolute value of the functions decreases
fast. When the plates separation is more than several times larger
than the extra dimensions radius, the absolute value will approach
to zero. The manifestation of extra dimensions influence on the
Casimir force on the piston under the limiting
$L\rightarrow\infty$ appears only when the distance between one
plate and the piston is about equal to the size of extra
dimensions. As mentioned above, if the extra dimensions possess
larger size, the extra-dimension corrections will appear clearly
in practice, therefore this model of one-dimensional piston can
become a window to examine the high-dimensional spacetime. It
should also be pointed out that the values of all correction
functions in the world with different dimensionality keep
negative, so the total Casimir force on the piston also remains
attractive. The experimental evidence shows that no repulsive
force generates in the case of parallel plates. It has been proved
theoretically and rigorously that the Casimir force between
parallel plates become repulsive as the plates are sufficiently
far away from each other in the higher-dimensional spacetime
described by Kaluza-Klein theory and the conclusion conflicts with
the experimental phenomena and is inevitable [28, 29]. In this
work our arguments on the piston in the same environment drawn
above under the limiting $L\rightarrow\infty$ is consistent with
the experimental phenomena. It is useful to consider the system
with a piston further. After one plate has been moved to the
remote place, the three-parallel-plate model can be thought as an
ordinary system consisting of two parallel plates in which one
plate is thought as a piston. Actually it is interesting that the
two kinds of results from three-parallel-plate model with limiting
$L\rightarrow\infty$ and original two-parallel-plate system
respectively are completely different. The theoretical finding in
this work, the Casimir force per unit area on the piston, is
consistent with the measurement at least qualitatively. The
deviation produces apparently as the plate-piston gap is close to
the extra-dimension size. Maybe the corrections are beyond the
experimental reach because the order of the compactification scale
of the additional spatial dimensions can be extremely tiny. The
three-parallel-plate model, called one-dimensional piston, must
replace the standard parallel plates system unless the
higher-dimensional approach described by Kaluza-Klein theory is
excluded.

In this work the model of three parallel plates in which the
middle one is called piston is studied in the higher-dimensional
spacetime described by Kaluza-Klein theory. The expression of
Casimir force per unit area on the piston is obtained. When one
outer plate is moved away, we also get the exact form of reduced
Casimir force per unit area between one plate and the piston. In
the limiting case we discover that the force is always attractive
and depend on the properties of extra compactified dimensions. The
more extra dimensions will produce stronger influence. When the
separation of one plate and the piston is larger enough than the
size of additional spatial dimensions and the limit
$L\rightarrow\infty$ is taken, the Casimir force per unit area
between them will be the same as the results for the standard
system consisting of two parallel plates in the four-dimensional
spacetime. The results of the standard system in four-dimensional
spacetimes are favoured in practice. Further we can argue that the
model of two parallel plates and one piston under the condition
that one outer plate has been moved extremely far away should
substitute the standard two-parallel-plate model to describe the
Casimir effect for parallel plates from experiment if the universe
has additional compactified dimensions. We also should not neglect
that the expressions of Casimir force for the two models derived
and calculated in the same high-dimensional spacetime and the
frame of the same Kaluza-Klein theory are completely different, in
which at least one is attractive and the other is repulsive.
Clearly in the high-dimensional background, our results from
plate-piston-plate model in the limiting case avoid the flaw of
results from general two-parallel-plate system. We should make use
of model of one-dimensional piston to explain the measurement of
Casimir effect from two paralle plates not a standard two-parallel
plate model. Of course the further consequences and related topics
are under progress.

\vspace{3cm}

\noindent\textbf{Acknowledgement}

The author thanks Professor K. Milton, Professor E. Elizalde and
Professor I. Brevik for helpful discussions. This work is
supported by the Shanghai Municipal Science and Technology
Commission No. 04dz05905 and NSFC No. 10333020.

\newpage
\begin{figure}
\setlength{\belowcaptionskip}{10pt} \centering
  \includegraphics[width=15cm]{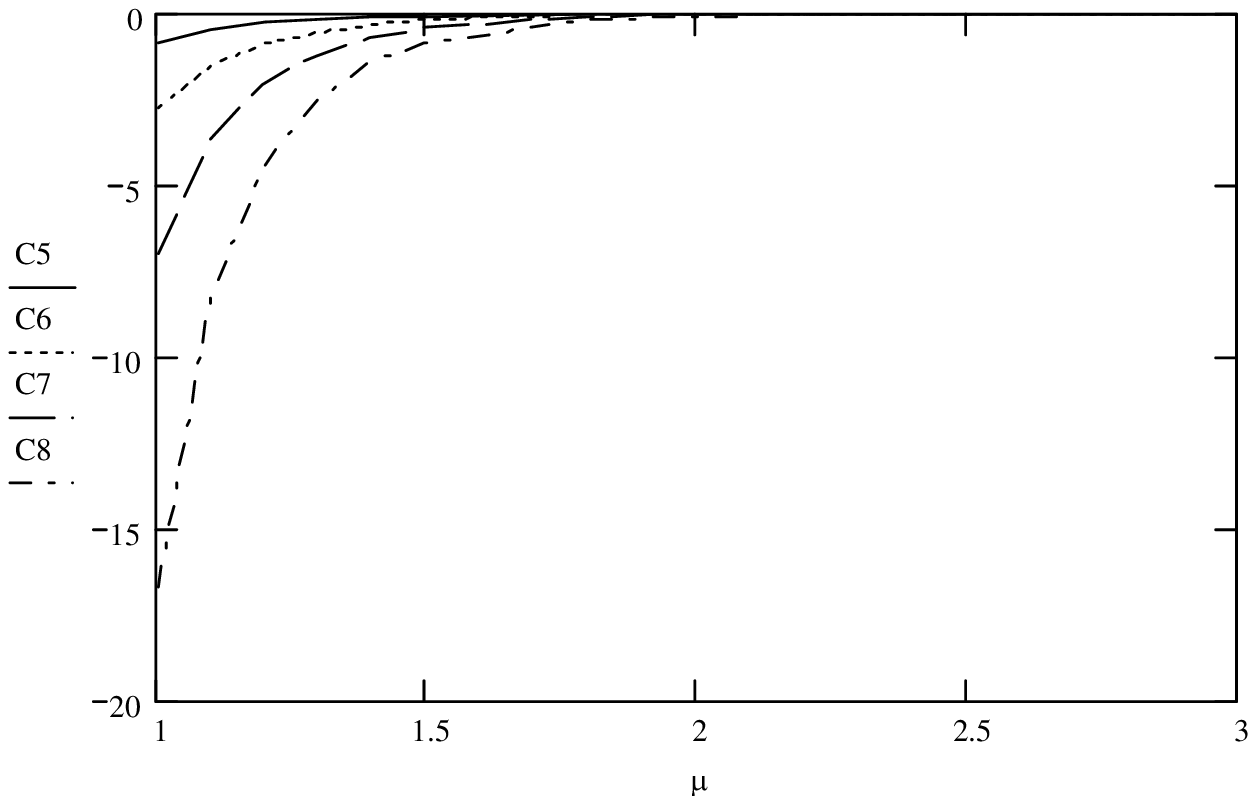}
  \caption{The solid, dot, dashed and dot-dashed curves of
  the correction functions of ratio of plate-piston distance
  and extra-dimension radius $\mu=\frac{a}{R}$ in ($4+d$)-dimensional spacetime for
  $d=1, 2, 3, 4$ respectively.}
\end{figure}

\begin{figure}
\setlength{\belowcaptionskip}{10pt} \centering
  \includegraphics[width=15cm]{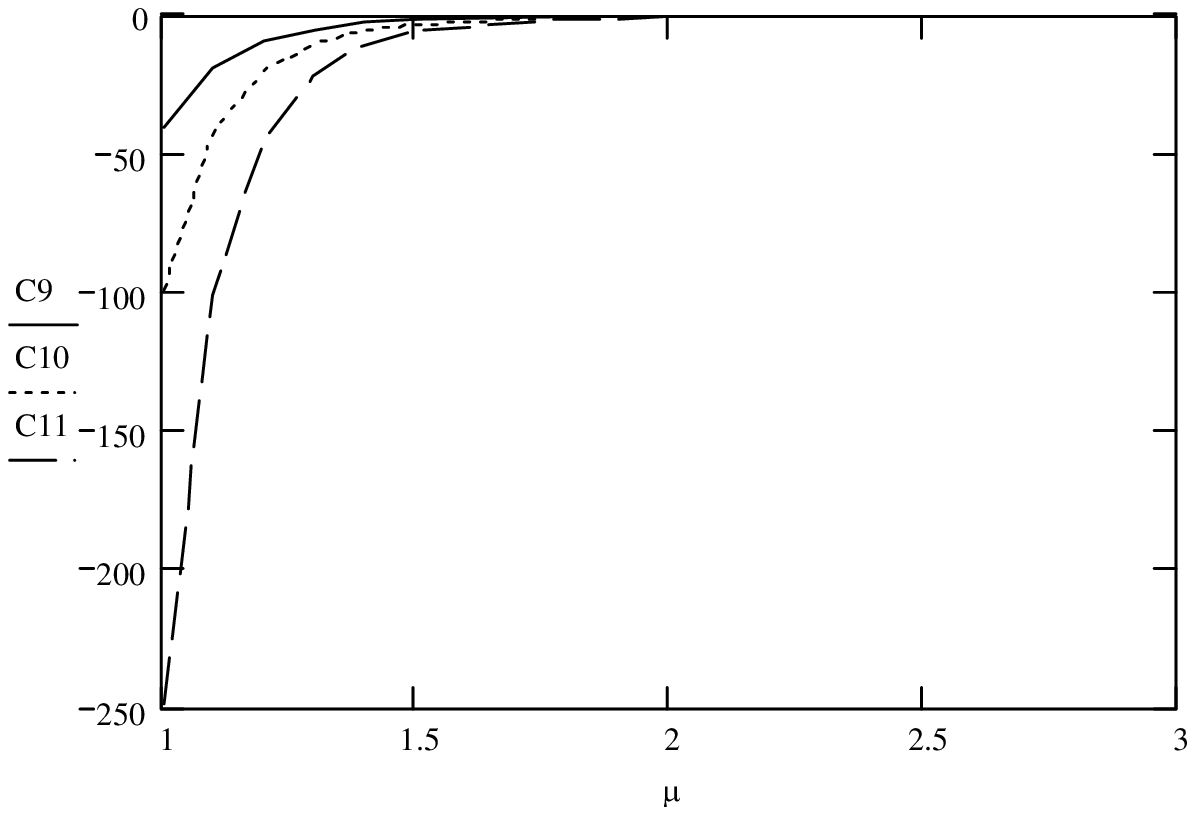}
  \caption{The solid, dot and dashed curves of the correction functions of
  ratio of plate-piston distance and extra-dimension radius $\mu=\frac{a}{R}$
  in ($4+d$)-dimensional spacetime for $d=5, 6, 7$ respectively.}
\end{figure}

\end{document}